\documentclass{pnastwo}

\usepackage{cuted}
\usepackage{upgreek}
\usepackage{cite}
\newlength\xparindent
    {%
        \par
        \hfill\rule[-6pt]{0.4pt}{6.4pt}%
        \rule{\dimexpr(0.5\textwidth-0.5\columnsep-1pt)}{0.4pt}
        \end{strip}
    }
\newcommand{\Bart}[1]{{\color{red}}} %For comments
\newcommand{\Romaric}[1]{{\color{blue}}} 
\newcommand{\Ricard}[1]{{\color{green}}} 

\newcommand{\rmd}{{\mathrm d}}

\renewcommand{\vec}[1]{\boldsymbol{#1}}

\newcommand*\colvec[3][]{
    \begin{pmatrix}\ifx\relax#1\relax\else#1\\\fi#2\\#3\end{pmatrix}
}

\usepackage{graphicx}
\usepackage[numbers,sort]{natbib}
\usepackage{amssymb,amsfonts,amsmath}
\usepackage[normalem]{ulem}
\DeclareGraphicsExtensions{.png,.pdf}

\url{www.pnas.org/cgi/doi/10.1073/pnas.0709640104}
\copyrightyear{2016}
\issuedate{December 20, 2016}
\volume{vol. 113}
\issuenumber{no. 51}

\begin{document}

%----------------------------------------------------------------------------------------
%	TITLE AND AUTHORS
%----------------------------------------------------------------------------------------

\title{Emergent structures and dynamics of cell colonies by contact inhibition of locomotion}

%------------------------------------------------

%% Enter authors via the \author command.  
%% Use \affil to define affiliations.
%% (Leave no spaces between author name and \affil command)

%% Note that the \thanks{} command has been disabled in favor of
%% a generic, reserved space for PNAS publication footnotes.

%% \author{\!<\!author name\!>\!
%% \affil{\!<\!number\!>\!}{\!<\!Institution\!>\!}} One number for each institution.
%% The same number should be used for authors that
%% are affiliated with the same institution, after the first time
%% only the number is needed, ie, \affil{number}{text}, \affil{number}{}
%% Then, before last author ...
%% \and
%% \author{\!<\!author name\!>\!
%% \affil{\!<\!number\!>\!}{}}

%% For example, assuming Garcia and Sonnery are both affiliated with
%% Universidad de Murcia:
%% \author{Roberta Graff\affil{1}{University of Cambridge, Cambridge,
%% United Kingdom},
%% Javier de Ruiz Garcia\affil{2}{Universidad de Murcia, Bioquimica y Biologia
%% Molecular, Murcia, Spain}, \and Franklin Sonnery\affil{2}{}}

\author{Bart Smeets\affil{1}{Division of Mechatronics, Biostatistics and Sensors (MeBioS) Kasteelpark Arenberg 30 - box 2456 3001 Leuven, Belgium},
Ricard Alert\affil{2}{Departament de F\'{i}sica de la Mat\`{e}ria Condensada \& Universitat de Barcelona Institute of Complex Systems (UBICS), Facultat de F\'{i}sica, Universitat de Barcelona, 08028 Barcelona, Spain}, 
Ji\v{r}\'{i} Pe\v{s}ek\affil{1}{},
Ignacio Pagonabarraga\affil{2}{},
Herman Ramon\affil{1}{}
\and
Romaric Vincent\affil{3}{Universit\'{e} Grenoble Alpes, Commissariat \`{a} l'\'{E}nergie Atomique (CEA) \& Laboratoire d'\'{E}lectronique des Technologies de l'Information (LETI), Micro and Nanotechnology Innovation Centre (MINATEC), F-38054 Grenoble, France}}

%\contributor{Submitted to Proceedings of the National Academy of Sciences of the United States of America}

%----------------------------------------------------------------------------------------

\significancetext{The regular distribution of mesenchymal cells, the formation of epithelial monolayers or their collapse into spheroidal tumors illustrate the broad range of possible organizations of cells in tissues. Unveiling a physical picture of their emergence and dynamics is of critical importance to understand tissue morphogenesis or cancer progression. Although the role of cell-substrate and cell-cell adhesion in the organization of cell colonies has been widely studied, the impact of the cell-type-specific contact inhibition of locomotion (CIL) remains unclear. Here, we include this interaction in simulations of active particles, and find a number of structures and collective dynamics that recapitulate existing tissue phenotypes. We give analytical predictions for the epithelial-mesenchymal transition and the formation of 3D aggregates as a function of cell-cell adhesion and CIL strengths. Thus, our findings shed light on the physical mechanisms underlying multicellular organization.}

\maketitle % The \maketitle command is necessary to build the title page

\begin{article}

%----------------------------------------------------------------------------------------
%	ABSTRACT, KEYWORDS AND ABBREVIATIONS
%----------------------------------------------------------------------------------------

\begin{abstract}
Cells in tissues can organize into a broad spectrum of structures according to their function. Drastic changes of organization, such as epithelial-mesenchymal transitions or the formation of spheroidal aggregates, are often associated either to tissue morphogenesis or to cancer progression. Here, we study the organization of cell colonies by means of simulations of self-propelled particles with generic cell-like interactions. The interplay between cell softness, cell-cell adhesion, and contact inhibition of locomotion (CIL) yields structures and collective dynamics observed in several existing tissue phenotypes. These include regular distributions of cells, dynamic cell clusters, gel-like networks, collectively migrating monolayers, and 3D aggregates. We give analytical predictions for transitions between non-cohesive, cohesive, and 3D cell arrangements. We explicitly show how CIL yields an effective repulsion that promotes cell dispersal, thereby hindering the formation of cohesive tissues. Yet, in continuous monolayers, CIL leads to collective cell motion, ensures tensile intercellular stresses, and opposes cell extrusion. Thus, our work highlights the prominent role of CIL in determining the emergent structures and dynamics of cell colonies.
\end{abstract}

%------------------------------------------------

 % When adding keywords, separate each term with a straight line: |

%------------------------------------------------

%% Optional for entering abbreviations, separate the abbreviation from
%% its definition with a comma, separate each pair with a semicolon:
%% for example:
%% \abbreviations{SAM, self-assembled monolayer; OTS,
%% octadecyltrichlorosilane}

% \abbreviations{}
% \abbreviations{CIL, Contact Inhibition of Locomotion; SPP, Self-Propelled Particles; MSD, Mean-Squared Displacement; EMT, Epithelial-Mesenchymal Transition}
%MIPS, Motility-Induced Phase Transition

%----------------------------------------------------------------------------------------
%	PUBLICATION CONTENT
%----------------------------------------------------------------------------------------

%% The first letter of the article should be drop cap: \dropcap{} e.g.,
%\dropcap{I}n this article we study the evolution of ''almost-sharp'' fronts

% \section{Introduction}
\dropcap{C}ell colonies exhibit a broad range of phenotypes. In terms of structure, collections of cells can arrange into distributions of single cells, assemble into continuous monolayers or multi-layered tissues, or even form 3D agglomerates. In terms of dynamics, cell motility may be simply absent, or produce random, directed or collective migration of cells. Transitions between these states of tissue organization are characteristic of morphogenetic events and are also central to tumor formation and dispersal \cite{Friedl2003,Friedl2009,Thiery2009,Nieto2013}. Therefore, a physical understanding of the collective behavior of cell colonies will shed light on the regulation of many multicellular processes involved in development and disease.

However, a complete physical picture of multicellular organization is not yet available, partly due to the challenge of modeling the complex interactions between cells. Here, we address this problem by means of large-scale simulations of self-propelled particles (SPP) endowed with interactions capturing generic cellular behaviors. Models of SPP with aligning interactions have been used to investigate collective cell motions in tissue monolayers
\cite{Mehes2014,Szabo2016,Szabo2006,Belmonte2008,Henkes2011,Basan2013,Sepulveda2013,Deforet2014,Woods2014,Tarle2015,Mones2015,Garcia2015,Zimmermann2016,Camley2016}. 
We extend this approach to unveil how the different structures and collective dynamics of cell colonies emerge from cell-cell interactions.

In addition to an excluded-volume repulsion, cells generally feature a short-range attraction as a consequence of their active cortical contractility transmitted through cell-cell junctions. With no additional interactions, this attraction would typically lead to cohesive tissues. However, not all cell types form cohesive tissues. Whereas epithelial cells tend to form continuous monolayers, mesenchymal cells separate after division despite the presence of cell-cell junctions. This observation calls for an extra effective repulsion to drive the separation, which may ultimately have a deep impact on the overall organization of the colony.

Such a repulsive interaction mediated by adhesion is indeed present in many cell types upon cell-cell contact, and is known as contact inhibition of locomotion (CIL) after Abercrombie and Heaysman \cite{Abercrombie1954}. Upon a cell-cell collision, the cell front adheres to the colliding cell, which hinders further cell protrusions. Subsequently, repolarization of the cell's cytoskeleton creates a new front away from the adhesion zone, and the two cells thus separate \cite{Abercrombie1958,Davis2015}. This interaction has been shown to be crucial in determining the collective behavior of cell groups in several contexts \cite{Desai2013}. For example, CIL guides the directional migration of neural crest cells \cite{Carmona2008}, and also ensures the correct dispersion of Cajal-Retzius cells in the cerebral cortex \cite{Villar-Cervino2013} or of hemocytes in the embryo \cite{Davis2012}.

Here, we model cellular interactions by means of an attraction due to intercellular adhesion, and a soft repulsion associated to the reduction of cell-substrate adhesion area. In addition, CIL is modeled as an interaction orienting cell motility away from cell-cell contacts. We analytically show how CIL acts as an effective repulsive force that hinders the formation of cohesive cell monolayers or 3D tissues at increasing cell-cell adhesion. We then explicitly predict the transitions between non-cohesive, cohesive, and overlapped organizations of the colonies as a function of cell-cell adhesion and CIL strength. In simulations, we identify states with different structures and emergent dynamics, including ordered or dynamic arrangements of clusters, gel-like networks, active gas and polar liquid states, and 3D aggregates. The results may be interpreted in biological terms by associating each state to common phenotypes, namely grid-like distributions of mesenchymal cells, collectively migrating epithelial monolayers, and cellular spheroids. The soft character of the potential and CIL interactions are key in producing structures and collective behaviors observed in cell colonies. In particular, the former enables the formation of 3D tissues via cell extrusion. In turn, CIL gives rise to self-organized collective motion in continuous cell monolayers. In line with \cite{Zimmermann2016}, we find that this effective repulsion induces tensile stresses in cell monolayers.

\section{Model}

We model a 2D colony of cells as a suspension of overdamped self-propelled disks. Neglecting translational noise, the equation of motion of cell $i$ with position $\vec{x}_i$ and polarity $\vec{p}_i \!=\! (\cos\theta_i,\sin\theta_i)$ reads
\begin{equation}\label{eq:eom}
F_m \vec{p}_i = \gamma_s \dot{\vec{x}}_i + \sum_j^{\text{nn}}\left[{F^{cc}_{ij}}\vec{\hat{n}}_{ij} +\gamma\left(\dot{\vec{x}}_i - \dot{\vec{x}}_j\right)\right],
\end{equation}
for contacting nearest neighbor cells $j$, with $\vec{\hat{n}}_{ij} \!=\! (\vec{x}_j - \vec{x}_i)/d_{ij}$ and $d_{ij} \!=\! ||\vec{x}_j - \vec{x}_i||$. Here, $F_m$ is the magnitude of the cell self-propulsion force, and $\gamma_s$ and $\gamma$ are cell-substrate and cell-cell friction constants, respectively.

\begin{figure}[bth]
\hspace*{1.cm}\textbf{(a)}\hspace*{2.2cm}\textbf{(b)}\hspace*{2.6cm}\textbf{(c)}\\
\vspace{0.1cm}
\centering
\includegraphics[width=0.95\columnwidth]{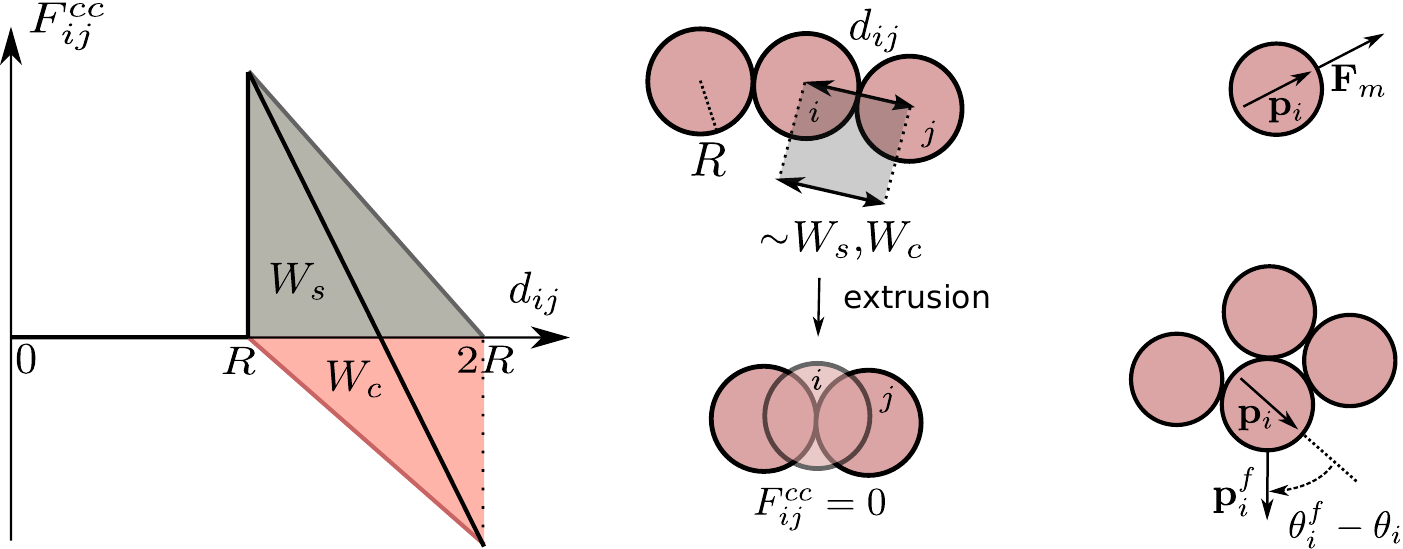}
\caption{\label{fig:model_scheme}A model of self-propelled particles with cell-like interactions. \textbf{(a)} Central cell-cell force $F^{cc}_{ij}$ (black) including a soft repulsion due to reduction of cell-substrate adhesion area (gray) and attraction due to active contractility through cell-cell adhesions (red). \textbf{(b)} Cell extrusion for intercellular distances $d_{ij}\!<\!R$, resulting in vanishing cell-cell forces in the plane. \textbf{(c)} Cellular self-propulsion force $F_m$ in the direction of the cell polarity $\vec{p}_i$. CIL rotates the polarity towards the direction $\vec{p}_i^f$ pointing away from cell-cell contacts.}
\end{figure}

The central force $F^{cc}_{ij}$ includes a soft repulsion $F^r_{ij}$ associated to the reduction of the cell-substrate adhesion area when two cells are closer than their spread size $2R$, and an attractive force $F^a_{ij}$ that accounts for active contractility transmitted through cell-cell adhesions. $F^r_{ij}$ is assumed to increase linearly with decreasing intercellular distance $d_{ij}$ up to $d_{ij}$\!=\!$R$. Hence, $F^r_{ij}\!=\!2W_s/R^2\left(2R-d_{ij}\right)$, with $W_s\!=\!\int_R^{2R} F^r_{ij}\,\rmd d_{ij}$ the cell-substrate adhesion energy (gray in Fig.\ \ref{fig:model_scheme}a). No further reduction of the cell-substrate contact area is allowed for $d_{ij}\!<\!R$. As a result, cells can approach at smaller distances under compression. In this regime cells do not exert any force on the substrate and are considered to be extruded from the monolayer (Fig.\ \ref{fig:model_scheme}a-b). Cell extrusions may lead to 3D tissues, whose structure and dynamics are not described by our 2D model.  $F^a_{ij}$ is assumed to increase linearly with distance up to $d_{ij}\!=\!2R$. Hence, $F^a_{ij}\!=\!-2W_c/R^2\left(d_{ij}-R\right)$, with $W_c\!=\!\int_R^{2R} F^a_{ij}\,\rmd d_{ij}$ the cell-cell adhesion energy (red in Fig.\ \ref{fig:model_scheme}a). Accordingly, the total interaction force (black in Fig.\ \ref{fig:model_scheme}a) reads
\begin{equation} \label{eq:force}
F^{cc}_{ij}\left(d_{ij}\right) = \begin{cases}
               \frac{2}{R}[W_s - \frac{W_s+W_c}{R}(d_{ij}-R)],& \text{if } R \leq d_{ij} \leq 2R\vspace{0.2cm}\\
               0, \ & \text{else.}
              \end{cases}
\end{equation}

In turn, CIL tends to orient the cell polarity $\vec{p}_i$ in the direction $\vec{p}_i^f$ pointing away from the weighted average position of the contacting cells (Fig.\ \ref{fig:model_scheme}c and SI Appendix). Similarly to \cite{Camley2016}, we model this interaction via a harmonic potential for the polarization angle $\theta_i$ that, in addition to rotational noise, yields
\begin{equation}\label{eq:angle_equation_arw}
 \dot{\theta_i} \!=\! - f_{\text{cil}} (\theta_i - \theta_i^f) + \sqrt{2 D_r} \,\xi.
\end{equation}
Here, $f_{\text{cil}}$ is the cellular repolarization rate upon cell-cell contact, whereas $\xi\left(t\right)$ is a typified Gaussian white noise, and $D_r$ is the rotational diffusion coefficient of cell motion.

The parameters of the model may be reduced to five dimensionless quantities: the packing fraction of cells $\phi$, cell-cell and cell-substrate adhesion energies $\overline{W}_c:=W_c/(2RF_m)$ and $\overline{W}_s:=W_s/(2RF_m)$, cell-cell friction $\overline{\gamma}:=\gamma/\gamma_s$, and a parameter $\psi:=f_{\text{cil}}/\left(2D_r\right)$ that compares the timescale of cytoskeletal repolarization associated to CIL to the rotational diffusion. Hereafter, we set $\phi\!=\!0.85$, $\overline{W}_s\!=\!1$, $\overline{\gamma}\!=\!0$, and focus on the effects of intercellular adhesion and CIL on the organization of cell colonies. The results are summarized in the phase diagram of Fig.\ \ref{fig:phase_diagram_colors}. Including cell-cell friction leads to jammed configurations of cohesive tissues (SI Appendix), in line with \cite{Garcia2015}. In turn, cell density does not affect the phase transitions but modifies the dynamical behavior of the cell colony (SI Appendix). Thus, cell proliferation may drive the colony through different dynamical states (SI Appendix).
%\vspace*{-0.25cm}
\section{Results}

\subsection{Non-cohesive phase}

\begin{figure}[tbh]
\centering
\includegraphics[width=0.85\columnwidth]{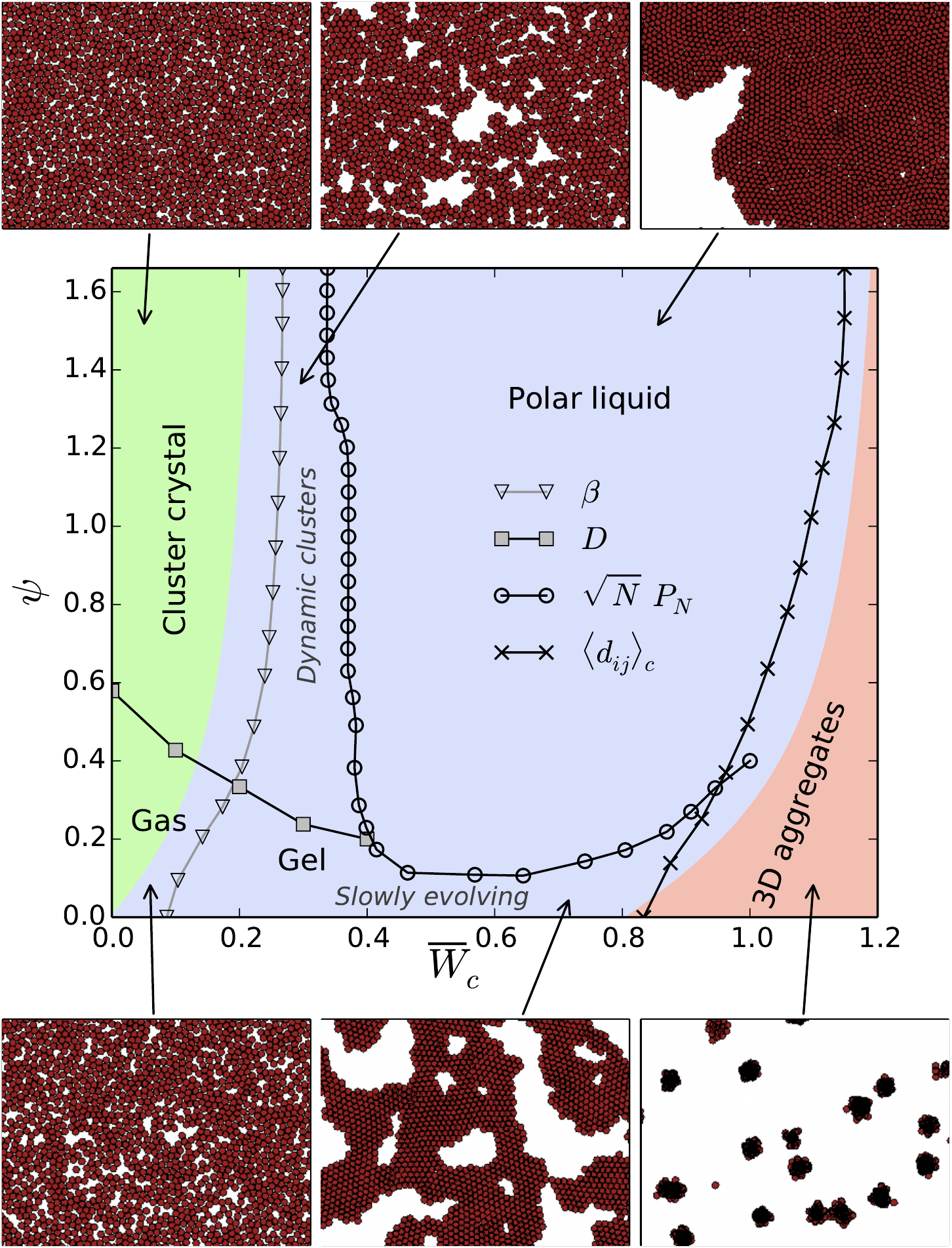}
\caption{\label{fig:phase_diagram_colors}Phase behavior of cell colonies as a function of cell-cell adhesion $\overline{W}_c$ and cell repolarization rate $\psi$ associated to CIL. Colors indicate the predicted regions for non-cohesive (green), cohesive (blue), and overlapped (red) organizations. In addition to capturing these structural transitions, simulations allow to identify dynamically distinct states such as an active gas, a cluster crystal, a gel-like percolated network, dynamic clusters, and an active polar liquid, as illustrated in snapshots.}
\end{figure}

We first study the transition between a cohesive phase in which cells remain in contact, $d_{ij}\!<\!2R$, and a non-cohesive phase in which they lose contact. Loss of cell contact is only possible if the maximal attractive force at $d_{ij}\!=\!2R$, $F^{cc}_{ij}\left(2R\right)\!=\!-2W_c/R$, is overcome by the component of the cells' self-propulsion force along the interparticle axis. Such component depends on the relative alignment of self-propulsion forces, and hence on CIL. When averaged over orientations, self-propulsion forces yield an effective central repulsion $\vec{F}^p_{ij}\!=\!\left\langle F_m\vec{p}_i\right\rangle_{\theta_i}$ between cells that depends on their repolarization rate $\psi$ (SI Appendix). In the relevant limit $\psi\gg 1/\left(2\pi\right)$ (Discussion), it reads
%\vspace*{-0.2cm}
\begin{equation} \label{eq:effective_repulsion}
\vec{F}^p_{ij}=\left\langle F_m\vec{p}_i\right\rangle_{\theta_i}\approx F_m \exp\left(-\frac{1}{4\psi}\right)\vec{\hat{n}}_{ij}.
\end{equation}
Then, within this mean-field approximation, the condition $F^p_{ij}+F^{cc}_{ij}\left(2R\right)\!=\!0$ gives a prediction for the transition between the non-cohesive (green in Fig.\ \ref{fig:phase_diagram_colors}) and cohesive (blue in Fig.\ \ref{fig:phase_diagram_colors}) phases. This sets a critical adhesion energy
%\vspace*{-0.2cm}
\begin{equation}\label{eq:wccoh}
\overline{W}_c^{\text{coh}}=\frac{1}{4} \exp\left(-\frac{1}{4\psi}\right),
\end{equation}
above which cells are expected to be in contact or, alternatively, a critical CIL rate above which cohesiveness is lost. Therefore, at low cell-cell adhesion, CIL promotes cell dispersal, thereby hindering the formation of cohesive tissues.

\begin{figure}[bth]
 \vskip -0.2cm
  \centering
  \hspace*{0.5cm}\textbf{(a)}\hspace*{3.9cm}\textbf{(b)}\\  
  \hspace*{0.1cm}\includegraphics[height=0.192\textheight]{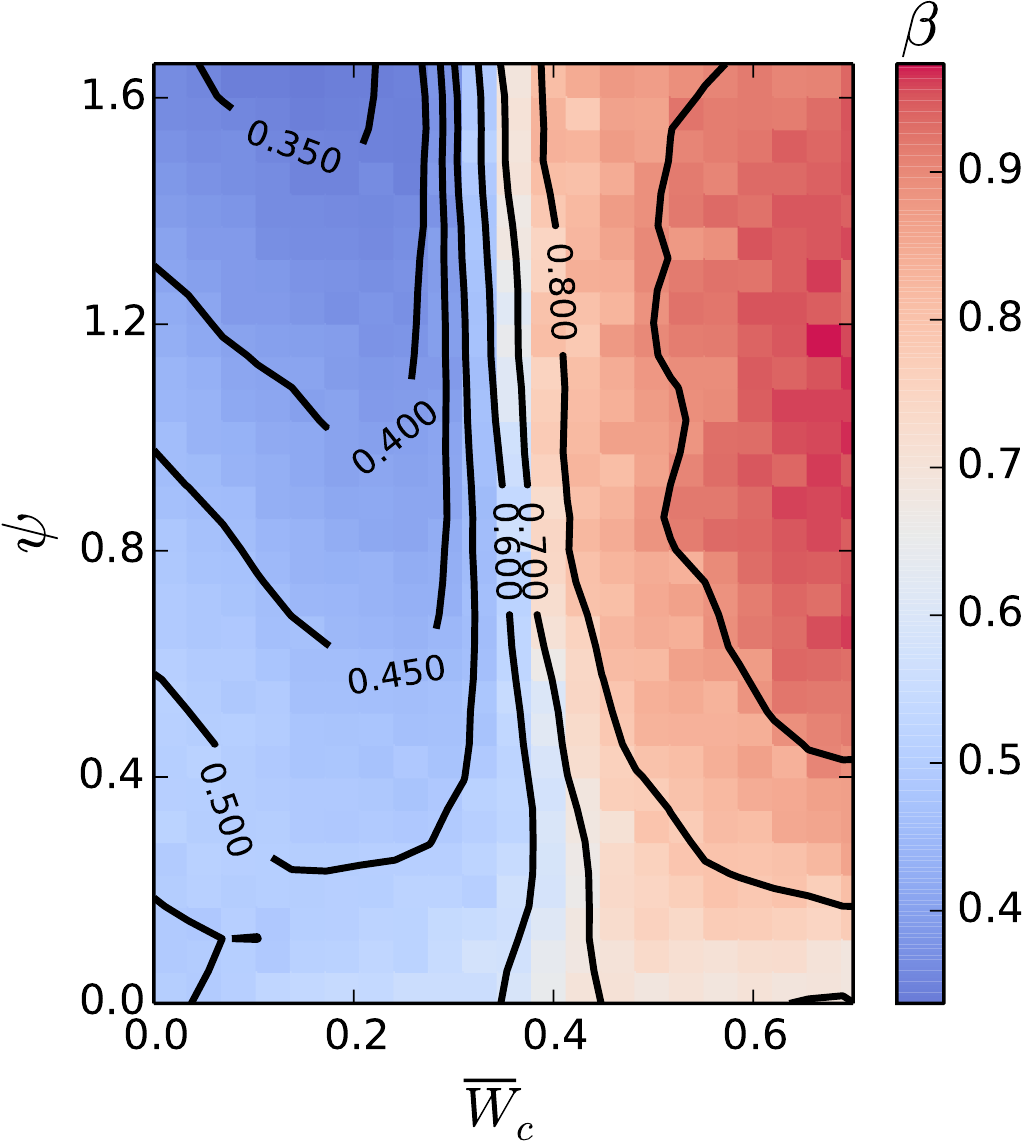}
  \hspace*{0.0cm}\includegraphics[height=0.182\textheight]{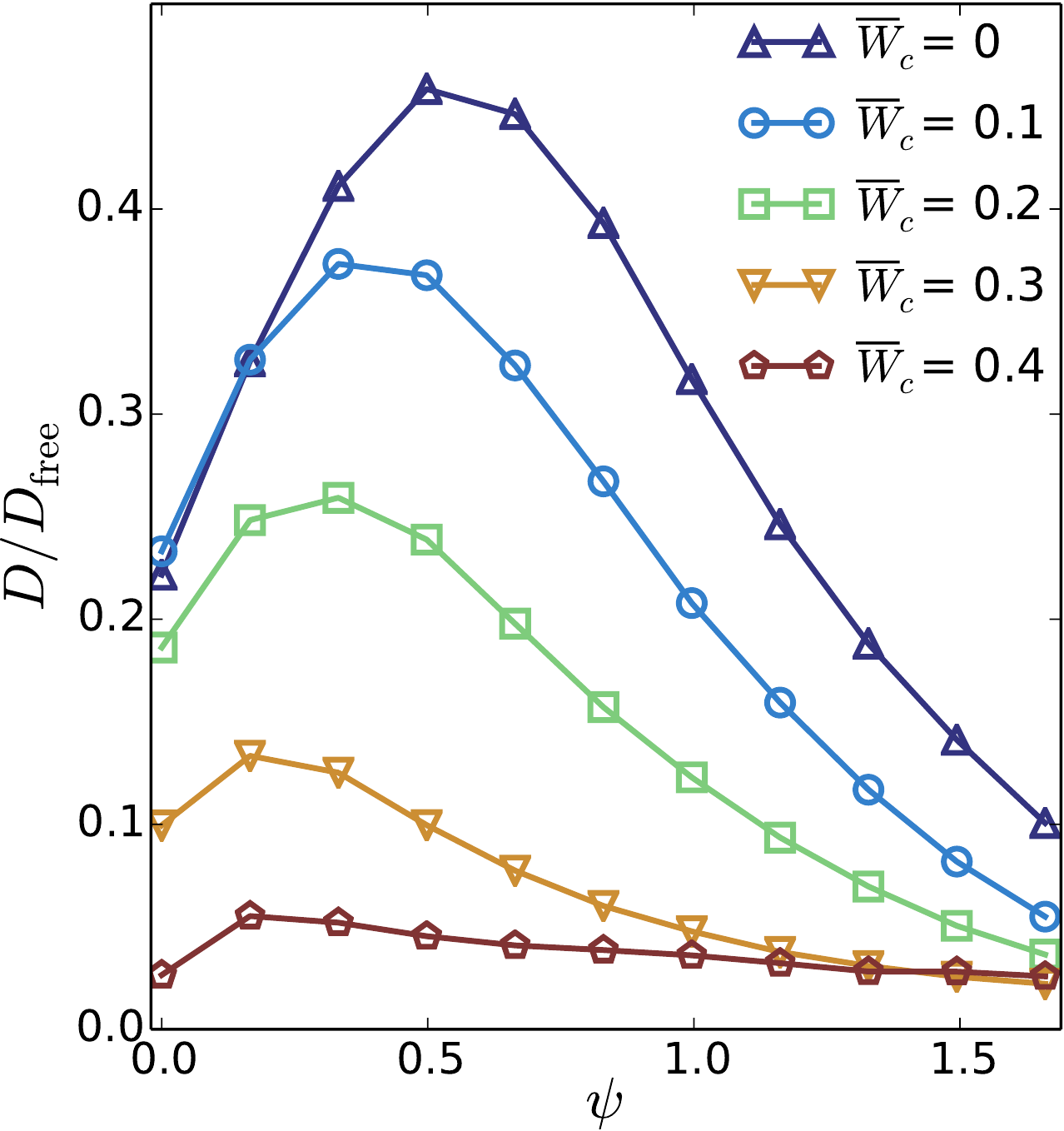}
  \caption{\label{fig:low_wc}Number fluctuations and diffusion in cell colonies. \textbf{(a)} Exponent of number fluctuations $\sigma_N\!\sim\! N^\beta$ as a function of cell-cell adhesion $\overline{W}_c$ and CIL repolarization rate $\psi$. Phase-separated states feature giant number fluctuations ($\beta\!>\!1/2$) whose onset identifies the transition to the cohesive phase (triangles in Fig.\ \ref{fig:phase_diagram_colors}). In the non-cohesive phase, colonies of slowly repolarizing cells (low $\psi$) feature equilibrium-like fluctuations ($\beta\!\approx\! 1/2$), whereas faster repolarizations (higher $\psi$) induce a hyperuniform distribution of cells ($\beta\!<\!1/2$). \textbf{(b)} Cell diffusion coefficient $D$ as a function of $\psi$ for some values of $\overline{W}_c$. For increasing repolarization rate $\psi$, $D$ initially increases but then decreases as clusters form. The maximum of $D\left(\psi\right)$ identifies the onset of clustering (squares in Fig.\ \ref{fig:phase_diagram_colors}). $D_{\text{free}}\!=\!F_m/(2\gamma_s D_r)$ is the translational diffusion coefficient of a persistent random walker with rotational diffusion \cite{Coffey2004}.}
\end{figure}

In simulations, we quantify this transition in terms of particle number fluctuations. Phase separated self-propelled disks feature giant number fluctuations \cite{Henkes2011,Fily2012,Fily2014}. There, the standard deviation of the number of particles $N$ in a given region scales as $\sigma_N\!\sim\! N^\beta$ for large $N$, with $\beta\!\approx\! 1$, whereas a system at equilibrium would feature $\beta\!=\!1/2$. Similarly, we compute the exponent $\beta$ (Fig.\ \ref{fig:low_wc}a) and identify the regions with $\beta\!>\!1/2$ as phase-separated, and thus cohesive. Consequently, we identify the transition to the cohesive phase from the onset of giant number fluctuations (triangles in Fig.\ \ref{fig:phase_diagram_colors}), which qualitatively agrees with the mean-field analytical prediction.

Within the non-cohesive phase (green in Fig.\ \ref{fig:phase_diagram_colors}), the colony forms an active gas state with equilibrium-like statistics ($\beta\!\approx\! 1/2$) at low CIL repolarization rates $\psi$ (Movie S1). At larger $\psi$, cells get hyperuniformly distributed, with $\beta\!<\!1/2$ (Fig.\ \ref{fig:low_wc}a), forming a crystal of small cell clusters (Movie S2). This state is reminiscent of the equilibrium cluster crystals formed by purely repulsive soft spheres \cite{Mladek2006}. In our case, an effective repulsion arises from anti-aligned propulsion forces via CIL (Eq.\ \ref{eq:effective_repulsion}). Similarly to \cite{Levis2014}, we set a dynamical criterion for the clustering transition based on the cell diffusion coefficient $D$ obtained from the long-time mean-squared displacement (MSD), $\lim_{t\rightarrow\infty}||\Delta\vec{x}||^2\!=\!4Dt$. Increasing the repolarization rate $\psi$ initially enhances diffusion by promoting cluster evaporation. However, the stronger effective repulsion at larger $\psi$ progressively prevents cells from escaping the clusters, hence reducing diffusion until it is eventually solely due to intercluster hopping events \cite{Moreno2007} (see SI Appendix for a discussion on the dependence of D on $\psi$). Consequently, we locate the clustering transition (squares in Fig.\ \ref{fig:phase_diagram_colors}) from the maximum of $D\left(\psi\right)$ at each $\overline{W}_c$ (Fig.\ \ref{fig:low_wc}b). Increasing cell-cell adhesion favors clustering, thereby enabling the short-range CIL-asso\-ci\-a\-ted repulsion responsible for the crystalline order.

\vspace*{-0.1cm}
\subsection{Cohesive phase}

\begin{figure}[tbp]
 \vskip -0.2cm
  \centering 
  \hspace*{0.3cm}\textbf{(a)}\hspace*{3.9cm}\textbf{(b)}\\
  \hspace*{0.1cm}\includegraphics[height=0.18\textheight]{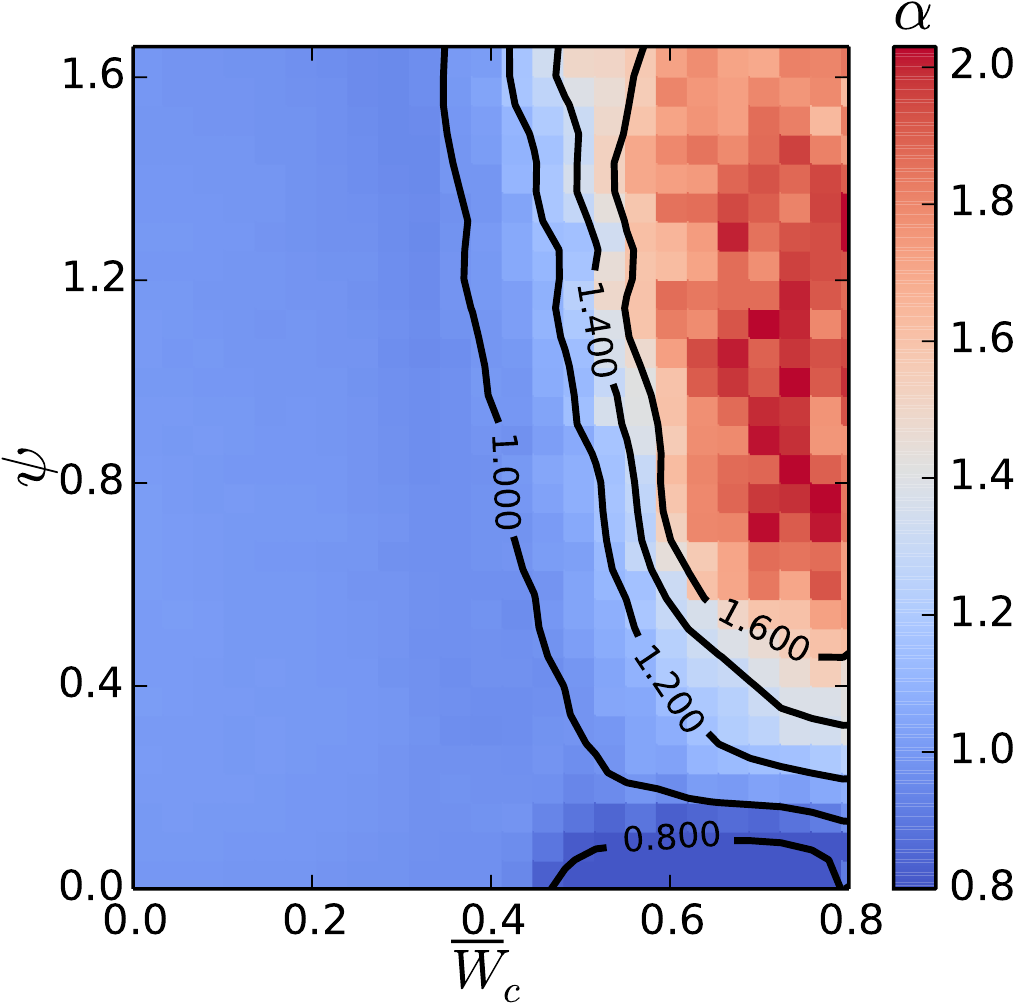}
  \hspace*{0.15cm}\includegraphics[height=0.18\textheight]{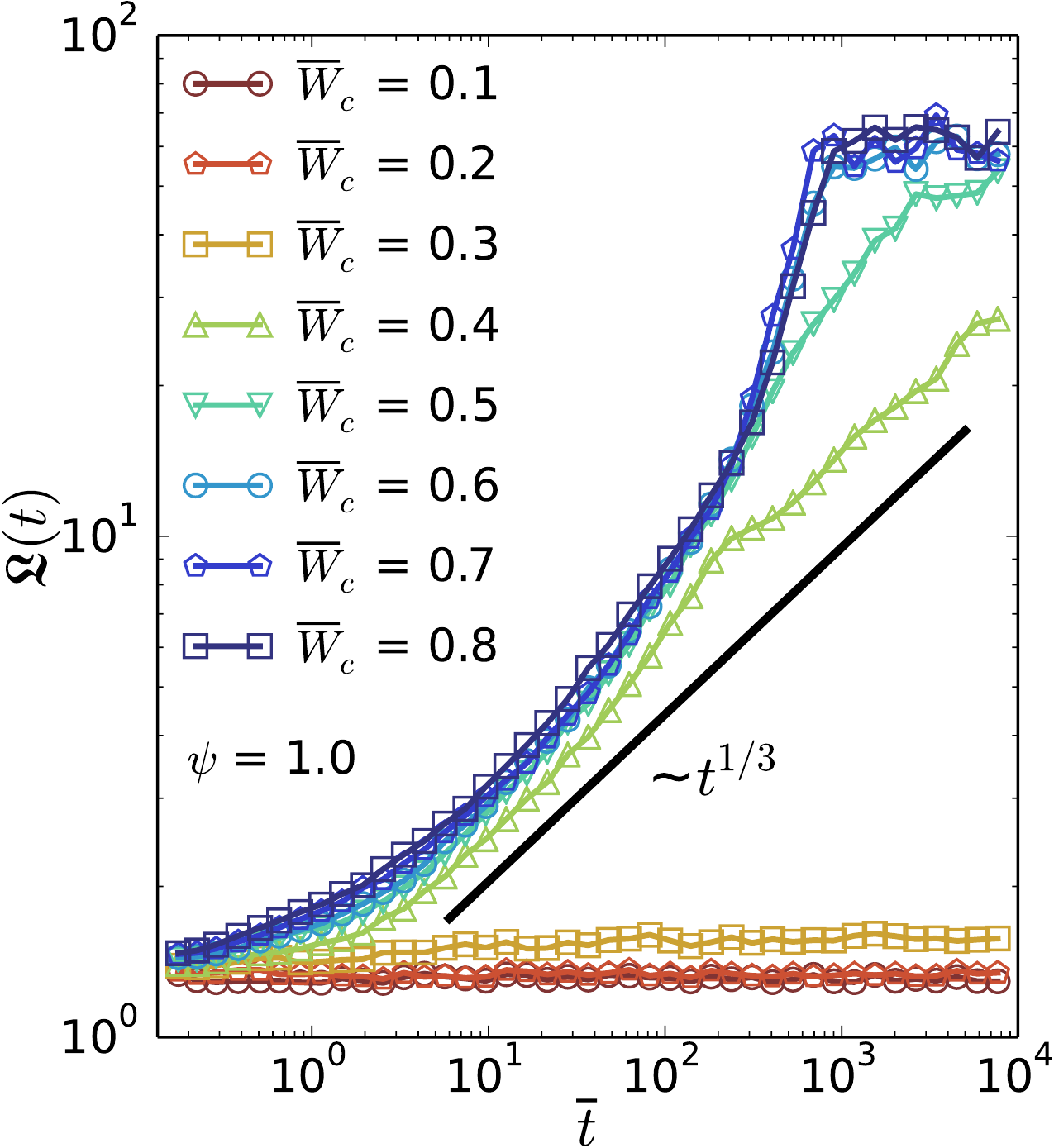}
  \textbf{(c)}
  \flushleft
  \footnotesize{\hspace*{0.5cm}$\overline{t}$ = 128\hspace*{1.1cm}320\hspace*{1.3cm}514\hspace*{1.3cm}704\hspace*{1.25cm}2000\hspace*{0.01cm}}\\
  \centering
  \includegraphics[width=0.47\textwidth]{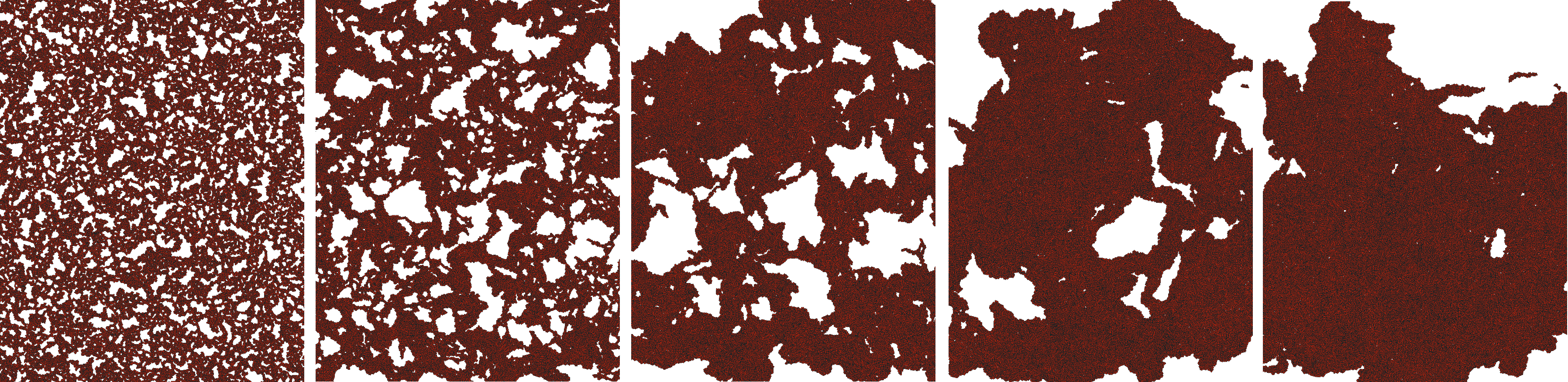}
      \vspace*{0.1cm}
  \caption{\label{fig:phs}Dynamics and phase-separation kinetics in cell colonies. \textbf{(a)} MSD Exponent $(\Delta\vec{x})^2\!\sim\! t^\alpha$ as a function of cell-cell adhesion $\overline{W}_c$ and CIL repolarization rate $\psi$. The colony forms a gel-like network with subdiffusive dynamics ($\alpha\!<\!1$) at low $\psi$. Faster CIL gives rise to collective cell motion as indicated by almost ballistic dynamics ($\alpha\!=\!2$). \textbf{(b)} Evolution of the average domain size $\mathfrak{L}\left(t\right)$, computed from the structure factor (SI Appendix), for different $\overline{W}_c$ at $\psi\!=\!1$. Dimensionless time reads $\overline{t}\!=\!F_m/(2R\gamma_s) t$. The colony phase separates for $\overline{W}_c\gtrsim 0.4$. CIL yields faster phase-separation kinetics than the diffusive coarsening dynamics of passive systems, for which $\mathfrak{L}\left(t\right)\!\sim\! t^{1/3}$ \cite{Bray2016}. \textbf{(c)} Illustration of the phase separation from an initial random configuration towards the active polar liquid at $\psi\!=\!1$ and $\overline{W}_c\!=\!0.7$.}
\end{figure}

Increasing cell-cell adhesion beyond the transition to the cohesive phase (blue in Fig.\ \ref{fig:phase_diagram_colors}), the colony initially forms a percolating structure of clusters. At low CIL repolarization rate $\psi$, cells arrange in a network with very slow, subdiffusive dynamics, as shown by the MSD $||\Delta\vec{x}||^2\!\sim\! t^\alpha$ with $\alpha\!<\!1$ (Fig.\ \ref{fig:phs}a). Thus, due to cell-cell adhesion, the colony forms a near-equilibrium attractive gel \cite{Redner2013b} with few cell rearrangements (Movie S3). At larger repolarization rates $\psi$ (above squares in Fig.\ \ref{fig:phase_diagram_colors}, see Fig.\ \ref{fig:low_wc}b), the effective CIL-associated repulsion yields smaller, dynamic, and locally crystalline clusters (Movie S4). They arise from a kinetic balance between the CIL-enhanced evaporation and the adhesion-induced condensation of clusters that prevents the completion of phase separation into a continuous dense phase.

Complete phase separation occurs at larger cell-cell adhesion, $\overline{W}_c\gtrsim 0.4$. The coarsening dynamics (Fig.\ \ref{fig:phs}b-c) are much faster than in a passive system, for which particle domains grow by diffusion as $\mathfrak{L}\left(t\right)\!\sim\! t^{1/3}$ \cite{Bray2016}. By orienting cell motility towards free space, CIL induces an advective coarsening of the cell domains that enables a fast phase separation of cell colonies.

Upon phase separation, the colony forms a continuous cell monolayer that exhibits self-organized collective motion (Movie S5). This is reflected in the MSD exponent, that evolves from diffusive ($\alpha\!=\!1$) towards almost ballistic ($\alpha\!=\!2$) above $\overline{W}_c\!\approx\! 0.4$ (Fig.\ \ref{fig:phs}a). CIL induces a coupling between cell polarity and density fluctuations in the fluid phase that gives rise to a macroscopic polarization via a spontaneous symmetry breaking. The outward motion of cells at the boundary of the monolayer creates free space behind them, which polarizes neighboring cells before the leading cell can reorient back. Through this mechanism, self-organized collective cell motion emerges from CIL, leading to an active polar liquid state.

The polar order is stable if the confinement imposed by neighbors restores the position and orientation of a cell before its polarity turns towards a new free direction. The repolarization occurs within a time-scale $1/f_{\text{cil}}$, and the characteristic time of position relaxation in a dense environment is $\sim\gamma_s/k$, with $k=4(W_s+W_c)/R^2$ the stiffness of a two-neighbor confinement. Thus, an approximate stability criterion reads $\gamma_s/k\lesssim f_{\text{cil}}^{-1}$, which is satisfied for the whole parameter range in Fig.\ \ref{fig:phase_diagram_colors} (SI Appendix).

As illustrated in Fig.\ \ref{fig:collective}a, isolated fluid monolayers may acquire a global polarity, and consequently perform persistent random walks with a persistence much larger than that of single cells (Movie S6). For randomly oriented cells, the average polarity of $N$ cells scales as $P_N \!=\! || \sum_{i=1}^N \vec{p}_i || / N \!\sim\! N^{-1/2}$. If cell polarities align, the average polarity of a small region of cells decreases slower with its size, so that $\sqrt{N}\,P_N\!>\!1$. The larger the repolarization rate $\psi$, the faster the increase of polarity with $N$ (SI Appendix). At sufficiently large sizes, multiple misaligned polarity domains appear that restore the random scaling (Fig.\ \ref{fig:collective}b). Hence, we define the onset of macroscopic polarization (circles in Fig.\ \ref{fig:phase_diagram_colors}) by the condition that $\sqrt{N}\,P_N$ has a maximum at $N\!=\!75$, namely that connected clusters consisting of up to $75$ cells may form a single polarity domain. The appropriate choice of $N$ depends on system size. However, for the sizes explored, the transition line (circles in Fig.\ \ref{fig:phase_diagram_colors}) is hardly sensitive to values around $N=75$ (SI Appendix). In conclusion, by ensuring a complete phase separation while still allowing for cell rearrangements, sufficiently strong cell-cell adhesion and CIL are required to form a polar, collectively moving cell monolayer.

Finally, the effective potential energy $E_p$ of cell-cell interactions gives information on the mechanics of the colony. Positive (negative) potential energies correspond to tensile (compressive) intercellular stresses. Non-cohesive colonies at low cell-cell adhesion feature average attractive interactions leading to the formation of clusters. In turn, by polarizing border cells outwards, CIL induces tensile stresses in cell monolayers (Fig.\ \ref{fig:collective}c), in agreement with \cite{Zimmermann2016}.

%\vspace*{-0.25cm}
\subsection{Overlapped phase}

We finally focus on the transition to 3D tissues. When the average total cell-cell force is attractive, cells eventually overcome the energy barrier associated to the soft repulsive potential (Fig.\ \ref{fig:model_scheme}a), which corresponds to cell extrusion events. Extruded cells are confined at distances smaller than $R$, where they exert neither cell-cell nor traction forces. Thus, our model can predict the onset of the transition to 3D cell arrangements. Assuming a homogeneous distribution of cells, and using Eq.\ \ref{eq:force}, the average interaction force reads
\begin{equation}
\left\langle F^{cc}_{ij}\right\rangle = \frac{\int_R^{2R} 2\pi d_{ij} F^{cc}_{ij}\;\rmd d_{ij}}{\int_R^{2R} 2\pi d_{ij}\;\rmd d_{ij}}=\frac{2}{9R}\left(4W_s-5W_c\right).
\end{equation}
This force adds to the effective repulsion $F^p_{ij}$ associated to anti-aligned self-propulsion forces (Eq.\ \ref{eq:effective_repulsion}), so that the transition between monolayers (blue in Fig.\ \ref{fig:phase_diagram_colors}) and 3D cell arrangements (red in Fig.\ \ref{fig:phase_diagram_colors}) is predicted by the condition $\left\langle F^{cc}_{ij}\right\rangle+F^p_{ij}\!=\!0$. This sets a critical cell-cell adhesion energy 
\begin{equation}\label{eq:wc3d}
\overline{W}_c^{\text{3D}}=\frac{1}{5}\left[4\overline{W}_s + \frac{9}{4} \exp{\left(-\frac{1}{4\psi}\right)}\right],
\end{equation}
above which cells are expected to fully overlap or, alternatively, a critical CIL repolarization rate above which cell extrusion is prevented. Therefore, by opposing cell extrusion, CIL hinders the collapse of cell monolayers into 3D aggregates. Indeed, a sufficiently fast repolarization of cell motility may stabilize cell monolayers even when cell-cell adhesion is stronger than cell-substrate adhesion, $\overline{W}_c\!>\!\overline{W}_s\!=\!1$ (Fig.\ \ref{fig:phase_diagram_colors}).

\begin{figure*}[btp]
\vskip -0.2cm
\centering
\textbf{(a)}\hspace*{3.9cm}\textbf{(b)}\hspace*{3.9cm}\textbf{(c)}\hspace*{3.9cm}\textbf{(d)}\\
    \includegraphics[height=0.18\textheight]{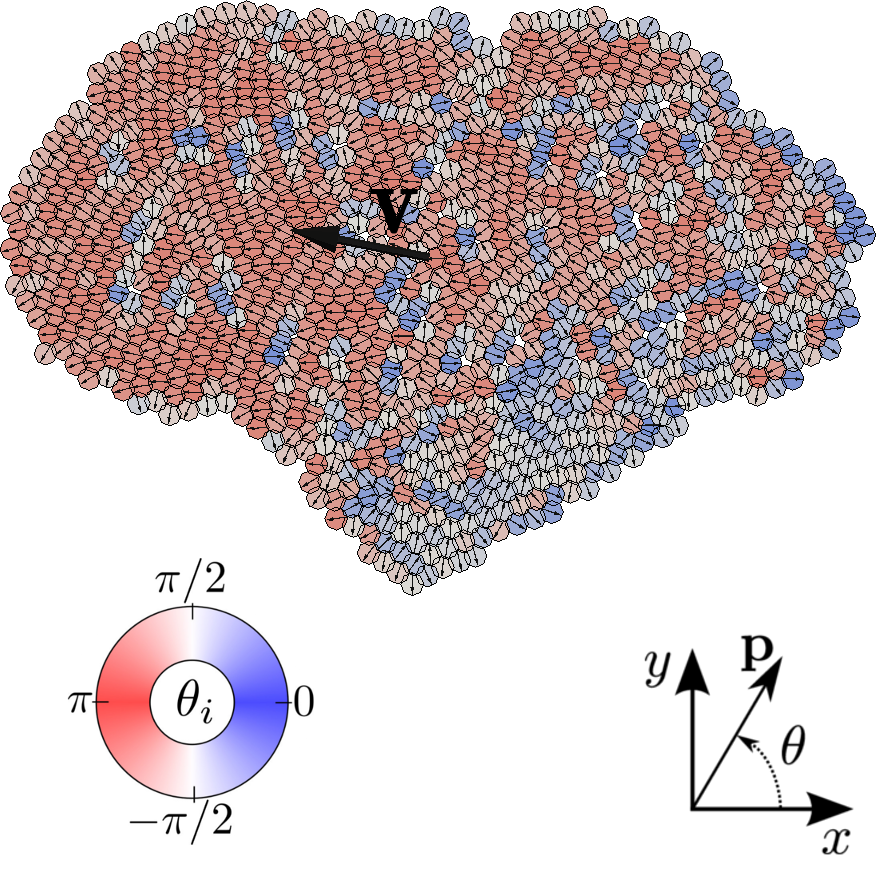}
     \hspace*{0.05cm}\includegraphics[height=0.18\textheight]{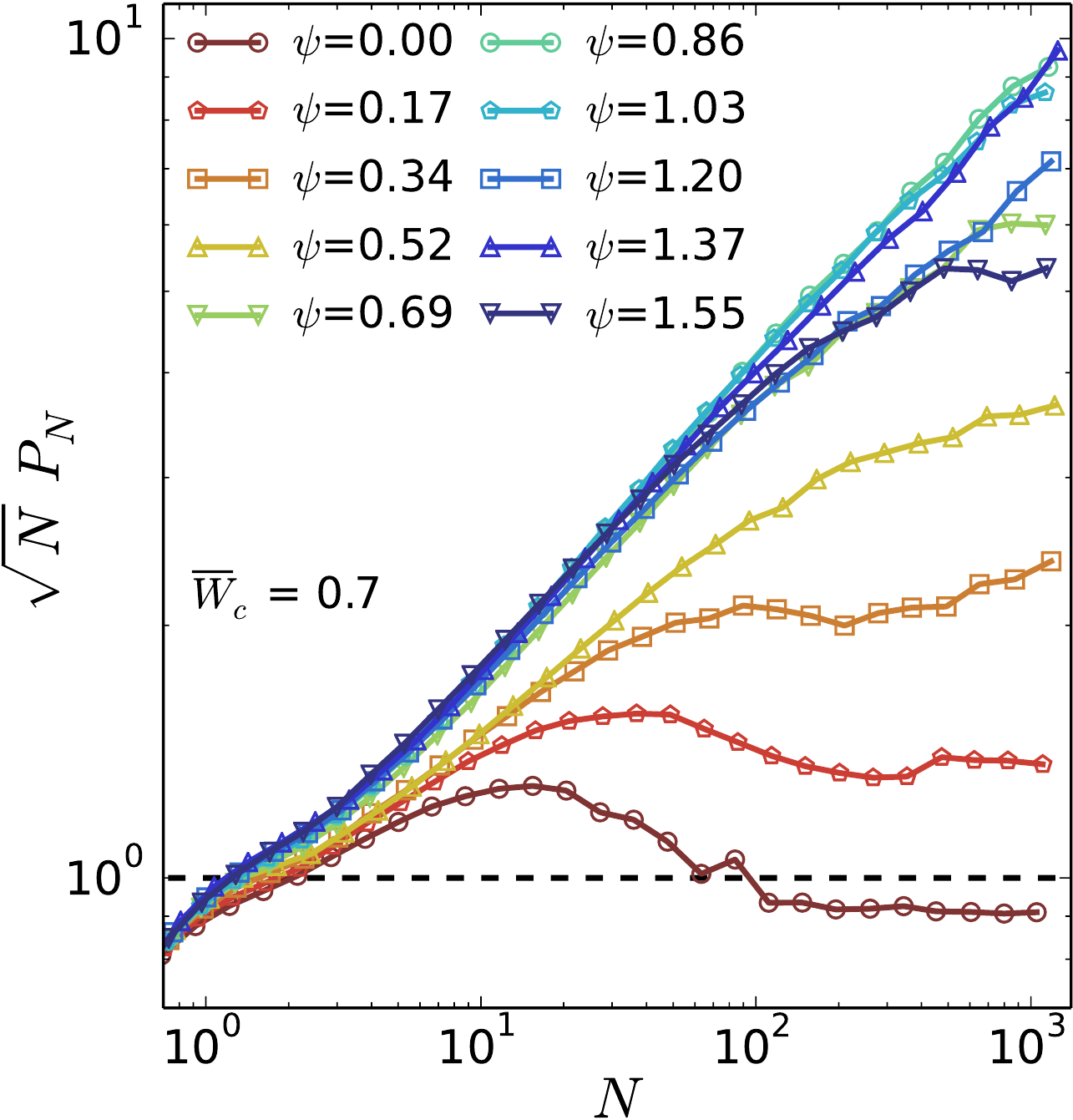}
     \hspace*{0.05cm}\includegraphics[height=0.19\textheight]{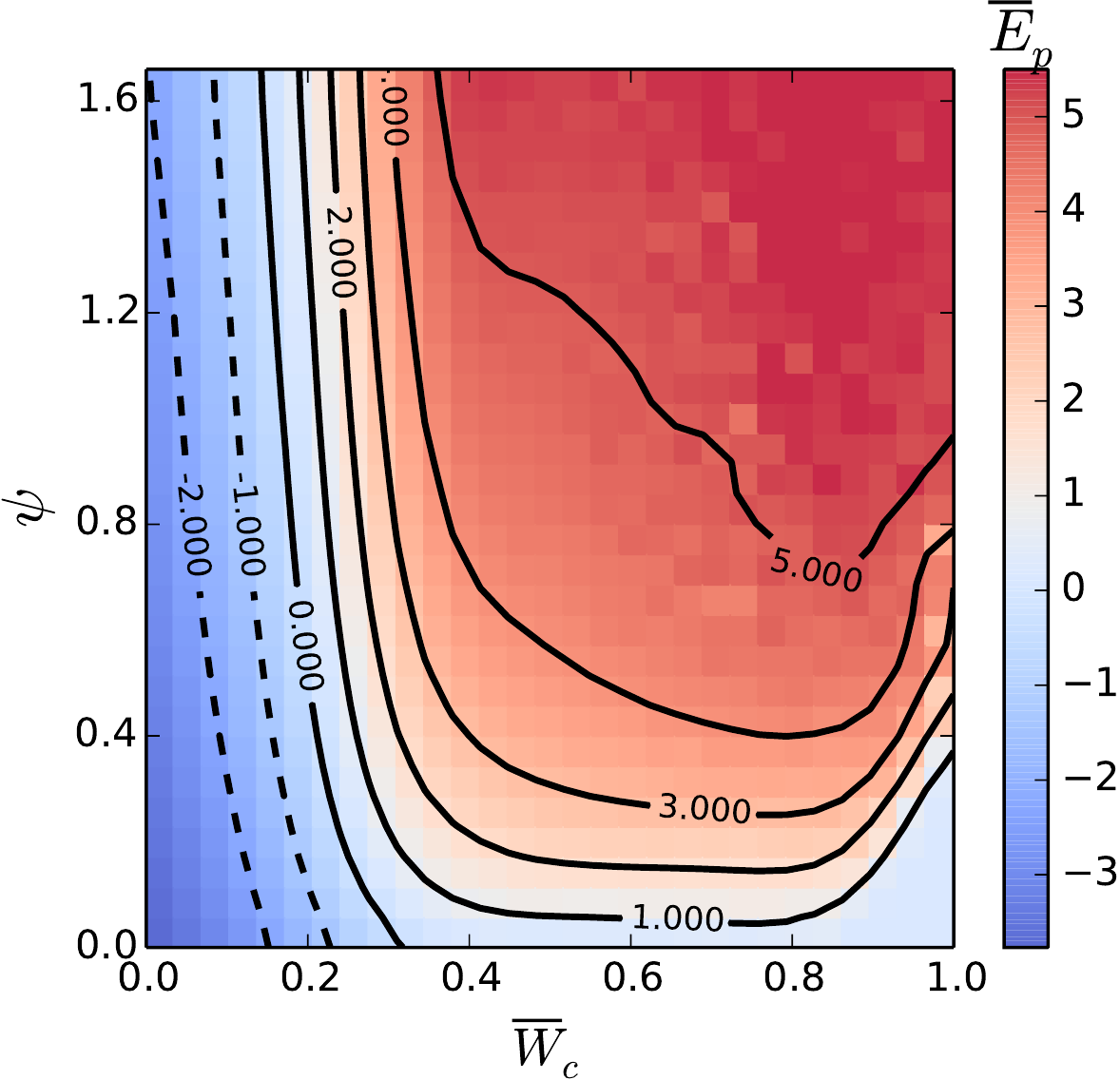}
     \hspace*{0.05cm}\includegraphics[width=0.17\textheight]{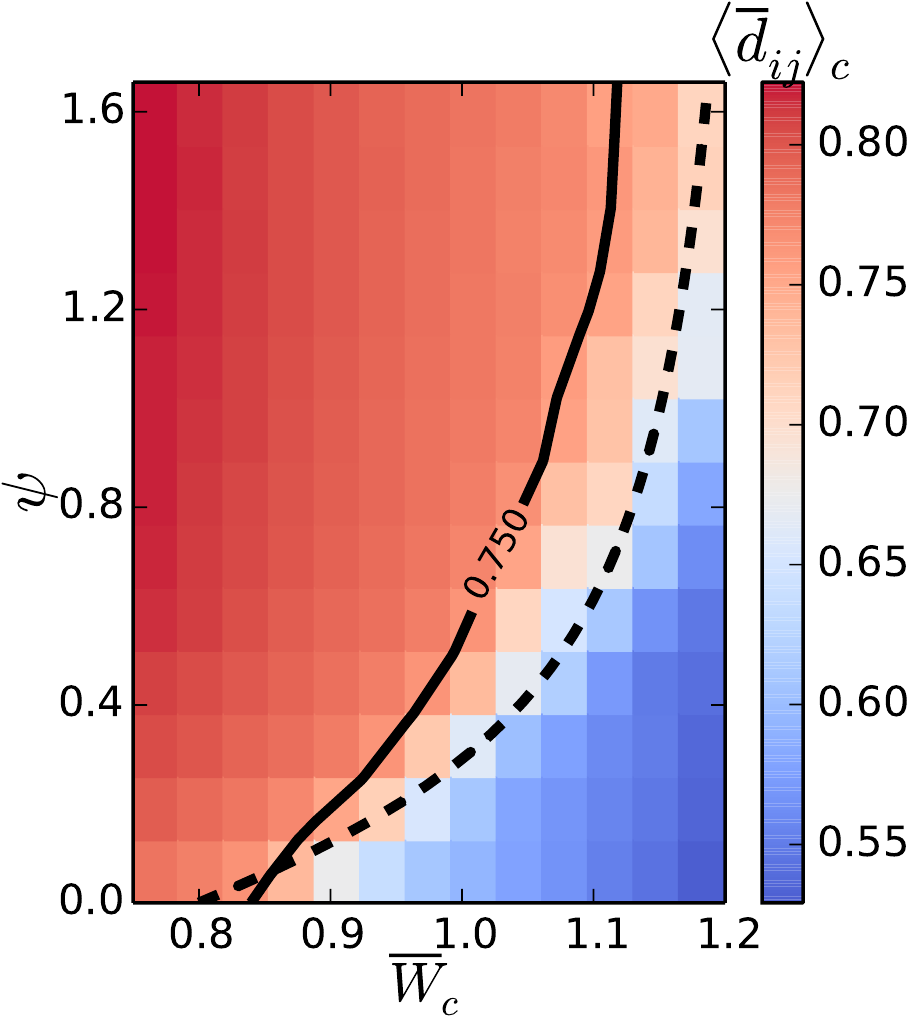}
\caption{\label{fig:collective}Collective motion, mechanics, and dewetting of cell monolayers. \textbf{(a)} Snapshot of a globally polarized, collectively migrating cell monolayer. \textbf{(b)} Rescaled average polarity $\sqrt{N}\,P_N$ of a monolayer of $N$ cells for different CIL repolarization rates $\psi$ at a cell-cell adhesion $\overline{W}_c\!=\!0.7$. $\sqrt{N}\,P_N\!=\!1$ corresponds to randomly oriented cells. CIL induces a global polarity ($\sqrt{N}\,P_N\!>\!1$) that gives rise to collective motion. The appearance of several polarity domains reduces the average polarity of large cell groups. The transition to the active polar liquid state (circles in Fig.\ \ref{fig:phase_diagram_colors}) is defined by the condition that the maximum of $\sqrt{N}\,P_N$ is at $N\!=\!75$. \textbf{(c)} Average cell-cell potential energy $\overline{E}_p\!=\!E_p/(2RF_m)$ as a function of cell-cell adhesion $\overline{W}_c$ and CIL repolarization rate $\psi$. CIL-associated repulsion induces tensile stresses ($\overline{E}_p\!>\!0$) in cell monolayers. \textbf{(d)} Average distance between contacting cells $\langle\overline{d}_{ij}\rangle_c\!=\!\left\langle d_{ij}\right\rangle_c/(2R)$ as a function of $\overline{W}_c$ and $\psi$. The transition between cell monolayers and 3D aggregates is predicted to occur at a vanishing average cell-cell force (dashed line), and identified by the condition $\left\langle d_{ij}\right\rangle_c\!=\!3R/2$ (solid line, crosses in Fig.\ \ref{fig:phase_diagram_colors}).\vspace{-0.25cm}}
\end{figure*}

In simulations, we characterize the degree of cell overlap in terms of the average distance between contacting cells $\left\langle d_{ij}\right\rangle_c$ (Fig.\ \ref{fig:collective}d). We then identify the transition when half of the contacting cells are at the critical distance for extrusion, $d_{ij}\!=\!R$, while the other half are fully spread, $d_{ij}\!=\!2R$. Hence, the transition is defined by $\left\langle d_{ij}\right\rangle_c\!=\!\frac{1}{2} R+\frac{1}{2}2R=3R/2$ (crosses in Fig.\ \ref{fig:phase_diagram_colors}), in qualitative agreement with the mean-field analytical prediction. 

Monolayer instability occurs through a dewetting process whereby holes appear in the cell monolayer, which rapidly evolves into a network structure, as observed in \cite{Douezan2012}. Subsequently, different regions of the network slowly collapse into separate aggregates (Movie S7). In general, the 3D aggregate - monolayer transition can be viewed as a wetting transition of the cell colony \cite{Douezan2011} enabled by cell insertion or extrusion \cite{Beaune2014}. Thus, our results show how CIL favors tissue wetting by orienting cell motility towards free space.

\vspace*{-0.5cm}
\section{Discussion and perspectives}

\begin{figure}[tbp]
\centering
\includegraphics[width=0.8\columnwidth]{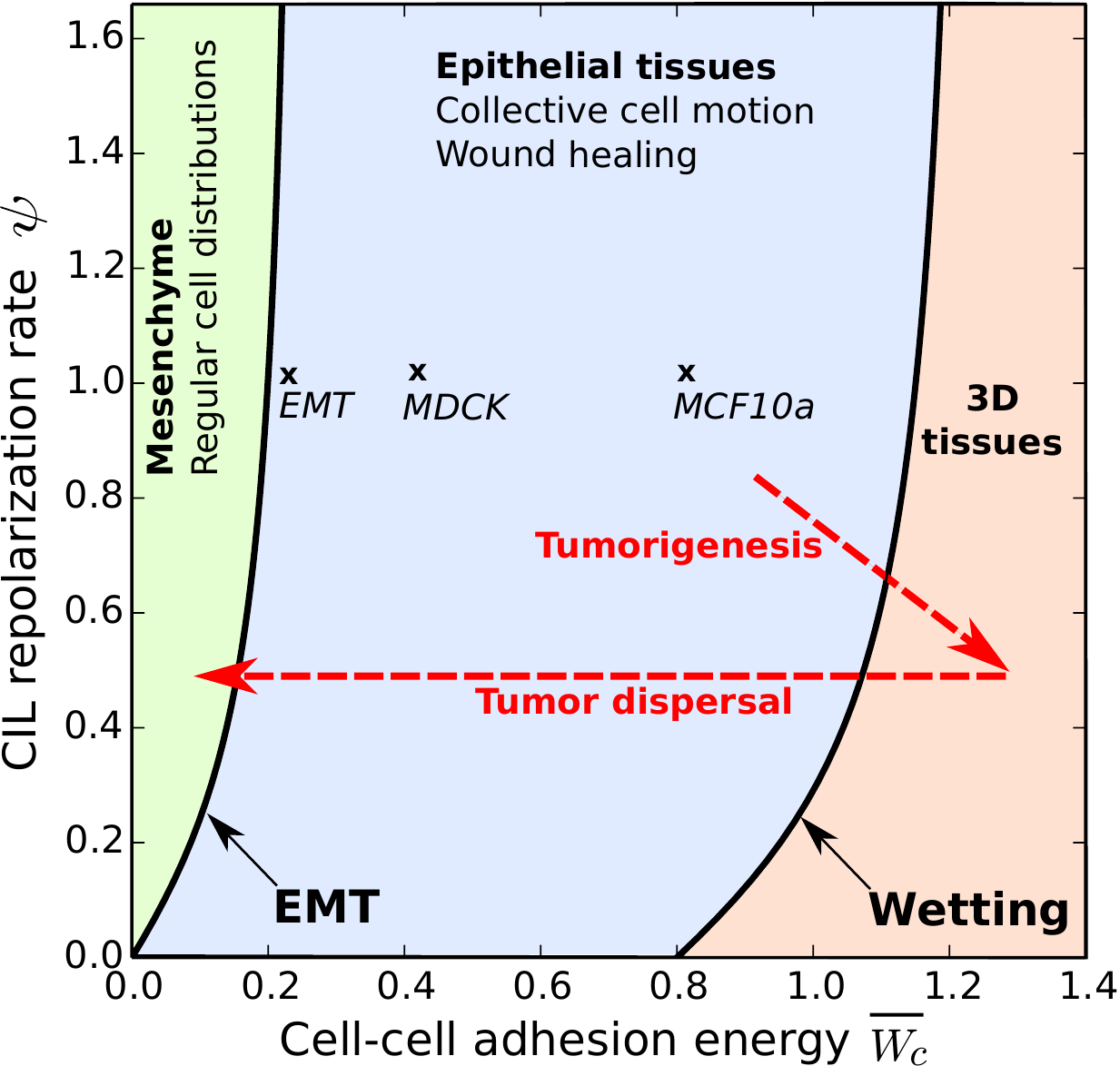}
\caption{\label{fig:exp_phase_diagram}Proposal for the classification of different tissue phenotypes (bold) in terms of the phases of the model (colors). The association is based on the indicated features, and supported by the parameter estimates for two epithelial tissues \cite{Vincent2015}, and an EMT \cite{Bazellieres2015} (crosses, see text). Speculated trajectories in cellular interaction parameters during cancer progression are also included (dashed arrows).}
\end{figure}

Based on experimental observations, we propose that the different organizations of cell colonies that emerge from our generic model correspond to different well-known tissue phenotypes (Fig.\ \ref{fig:exp_phase_diagram}). First, the non-cohesive phase, in which cells are not in contact, might correspond to mesenchymal tissues. Experiments show that CIL leads to regular distributions of mesenchymal cells during development \cite{Villar-Cervino2013,Davis2012}. This result is consistent with the transition towards an ordered structure of cell clusters by increasing CIL strength $\psi$ (Fig.\ \ref{fig:phase_diagram_colors}).

The cohesive phase, in which cells maintain contact, can correspond to epithelial tissues. In the active polar liquid state, CIL induces cells to spontaneously invade empty spaces within the tissue, similarly to wound healing processes characteristic of epithelia. Indeed, simulations of prepared wounds reproduce the closure dynamics observed in experiments \cite{Cochet-Escartin2014} (SI Appendix). In the absence of CIL, healing is severely impaired (SI Appendix), in agreement with experiments upon inhibition of Rac1 \cite{Anon2012}, a key protein for CIL behavior \cite{Roycroft2016}.

In addition, the parameters of our phase diagram can be estimated from experiments for two epithelial cell lines. By fitting the MSD of a SPP with rotational diffusion \cite{Coffey2004}, $||\Delta\vec{x}||^2\!=\!2v_m^2/D_r^2\left(D_rt+e^{-D_rt}-1\right)$, to experimental data for MCF10a cells (SI Appendix), we estimate a self-propulsion velocity $v_m\!=\!F_m/\gamma_s\!\approx\! 1$ $\upmu$m/min, and a diffusion coefficient $D_r\!\approx\! 0.05$ min$^{-1}$. This gives a P\'{e}clet number $\text{Pe}=3v_m/(2RD_r)\!\approx\! 2$, too low to produce motility-induced phase separation \cite{Redner2013a,Redner2013b}. In turn, the duration of cell-cell contact during CIL events allows the estimation of the rate of repolarization of cell motility. For two mesenchymal cell types, hemocytes \cite{Davis2015} and fibroblasts \cite{Kadir2011}, this gives $f_{\text{cil}}\!\approx\! 0.1$ min$^{-1}$. The same estimate is obtained for epithelial MDCK cells from the time that a wound needs to start closing \cite{Brugues2014}. Then, assuming these parameter values are similar for MCF10a and MDCK cells, we estimate $\psi:=f_{\text{cil}}/(2D_ r)\!\approx\! 1$ for both cell lines. Self-propulsion forces can be estimated from traction force measurements, which yield $F_m\!\approx\! 60$ nN and $F_m\!\approx\! 25$ nN in MCF10a and MDCK tissues, respectively \cite{Bazellieres2015}. Finally, cell-cell and cell-substrate adhesion energies can be related to an effective elastic modulus $\Gamma$ of an expanding monolayer, and to the total cellular strain $\epsilon_{\text{tot}}$ at which the expansion stops \cite{Vincent2015}. From Eq.\ \ref{eq:force}, $\Gamma\!\approx\!\left(W_s+W_c\right)/R^2$. In turn, $\epsilon_{\text{tot}}$ corresponds to the cell-cell distance at mechanical equilibrium, $d_{ij}^{\text{eq}}\!=\!\left(1+\epsilon_{\text{tot}}\right)R$, namely at which the total cell-cell force $F_{ij}^{cc}+F^p_{ij}\!=\!0$ vanishes:
%\vspace*{-0.25cm}
\begin{equation}
d_{ij}^{\text{eq}}=\frac{R}{W_s+W_c}\left[2W_s+W_c+\frac{RF_m}{2}\exp\left(-\frac{1}{4\psi}\right)\right].
\end{equation}
% \vspace*{-0.2cm}
Then, using $R\!=\!16$ $\upmu$m and the values of $\Gamma$ and $\epsilon_{\text{tot}}$ reported in \cite{Vincent2015}, we infer $\overline{W}_s\!\approx\! 1.1$ and $\overline{W}_c\!\approx\! 0.8$ for the MCF10a tissue, and $\overline{W}_s\!\approx\! 0.35$ and $\overline{W}_c\!\approx\! 0.42$ for the MDCK tissue. The transition to 3D structures would then occur at $\overline{W}_c^{3D}\!\approx\! 1.3$ for the MCF10a tissue. Thus, this tissue type falls well within the polar liquid state, in which cells form a collectively migrating continuous monolayer as experimentally observed. In contrast, the MDCK tissue is closer to the wetting transition, which we estimate at $\overline{W}_c^{3D}\!\approx\! 0.63$. Thus, although the latter also falls within the polar liquid state, it may form 3D structures more easily, in line with experimental observations \cite{Deforet2014}.

Now, the transition from cohesive to non-cohesive phases should correspond to the epithelial-mesenchymal transition (EMT), which is associated to down-regulation of cell-cell adhesion proteins \cite{Thiery2009,Nieto2013}. Our prediction sheds light on the role of CIL in the EMT (Fig.\ \ref{fig:exp_phase_diagram}). As above, we can estimate the parameters for an EMT in an expanding MCF10a monolayer. Upon a knockdown of cell-cell adhesion proteins, the epithelial tissue disaggregates at an intercellular stress $\sigma_{xx}^{\text{coh}}\!\approx\! 300$ Pa \cite{Bazellieres2015}. This translates into the critical cell-cell adhesion for the loss of cohesiveness by $\overline{W}_c^{\text{coh}}\!\approx\!\sigma_{xx}^{\text{coh}}hR^2$, with $h\!\approx\! 5$ $\upmu$m the height of the monolayer. Hence, we estimate $\overline{W}_c^{\text{coh}}\!\approx\! 0.2$, consistent with the prediction $\overline{W}_c^{\text{coh}}\!\approx\! 0.19$ at $\psi\!=\!1$.

A tissue may also undergo an EMT by increasing cell traction forces, such as upon treatment with hepatocyte growth factor \cite{Vincent2015,Maruthamuthu2014}. In our diagram, an increased self-propulsion force $F_m$ yields a lower dimensionless cell-cell adhesion energy $\overline{W}_c:=W_c/(2RF_m)$ whereas its critical value depends only on CIL (Eq.\ \ref{eq:wccoh}) hence causing the EMT.

In conclusion, the estimates and observations support the association of epithelial tissues to the cohesive phase. Nevertheless, some mesenchymal cells can also migrate collectively as a consequence of CIL \cite{Carmona2008,Woods2014,Szabo2016} or of increased cell-cell adhesion \cite{Plutoni2016}. Therefore, these specific phenotypes might also correspond to the active polar liquid state. However, whether the features of collective mesenchymal cell migration \cite{Theveneau2013} fully agree with our results deserves further exploration.

Finally, in our model, the overlapped phase corresponds to 3D tissues. Their structure is not captured by our 2D model, which only predicts the onset of their appearance. In experiments, the transition from a cell monolayer to a 3D aggregate can be induced in many ways \cite{Gonzalez2012}, such as by increasing the density of cell-cell adhesion proteins \cite{Ryan2001,Douezan2011}. Alternatively, one can reduce the density of cell-substrate proteins \cite{Ryan2001,Ravasio2015} which, in our diagram, entails a decrease of the critical cell-cell adhesion for the wetting transition, Eq.\ \ref{eq:wc3d}. 3D aggregates also form when the substrate is softened \cite{Douezan2012a}, which simultaneously decreases cell tractions $F_m$ and cell-substrate adhesion $W_s$. This increases $\overline{W}_c:=W_c/(2RF_m)$ while keeping $\overline{W}_s:=W_s/(2RF_m)$, and hence $\overline{W}_c^{\text{3D}}$, constant.

The monolayer-spheroid transition has been put forward as an in-vitro model for tumor formation and spreading \cite{Gonzalez2012}. In this context, our predictions may contribute to appreciate the role of CIL in cancer progression \cite{Abercrombie1979} (Fig.\ \ref{fig:exp_phase_diagram}). Indeed, downregulation of cell-cell adhesion and enhanced traction forces promote metastasis, which may proceed through many steps involving collective cell migration, dissemination of cell clusters, and a final EMT \cite{Friedl2003,Friedl2009,Thiery2009,Nieto2013,Cheung2016}.
%Our phase diagram may thus provide clues to unveil how modifications in cellular interactions drive the phenotypic evolution during cancer progression.

\vspace*{-0.4cm}
\section{Conclusions}
In summary, we studied the organization of cell colonies by means of self-propelled particle simulations. The interactions capture specific cellular behaviors such as CIL, and give rise to several structures and collective dynamics (Fig.\ \ref{fig:phase_diagram_colors}). Our results show how CIL leads to regular cell arrangements, and hinders the formation of cohesive tissues, as well as their extrusion-mediated collapse into 3D aggregates. Self-organized collective cell motion, with tensile intercellular stresses, also emerges from CIL interactions.

In addition, we have analytically derived an effective CIL-induced cellular repulsion force, which yields explicit predictions for transitions between non-cohesive, cohesive, and 3D colonies. Based on experimental observations and parameter estimates, we associate these phases to mesenchymal, epithelial, and 3D tissue phenotypes. Thus, our predictions may have implications for processes in development and disease that modify the tissue phenotype. In general, our active soft matter approach paves the way towards a physical understanding of multicellular organization and collective cell behavior.
%In future work, extending our model to 3D offers prospects of further insights.
\vspace*{-0.6cm}
\section{Methods}
We performed simulations of SPP in an overdamped system. Velocities are computed by solving $\vec{\underline{F}} \!=\! \underline{\underline{\vec{\Gamma}}}\cdot \vec{\dot{\underline{x}}}$. Positions are updated using an explicit Euler Scheme and the orientations using the Euler-Maruyama method, with $\Delta \overline{t}=$0.016. We simulate rectangular domains of 25$\times$10$^3$ up to 10$^5$ cells, enclosed by means of a stiff repulsive potential. To avoid boundary effects, cells close to the border are excluded from the analysis. A full description of the methods is given in the SI Appendix.

%----------------------------------------------------------------------------------------
%	APPENDICES (OPTIONAL)
%----------------------------------------------------------------------------------------

%----------------------------------------------------------------------------------------
%	ACKNOWLEDGEMENTS
%----------------------------------------------------------------------------------------

\vspace*{-0.2cm}
\begin{acknowledgments}
We thank E.\ Bazeli\`eres and X.\ Trepat for experimental help, and T.\ Odenthal and S.\ Vanmaercke for help with implementation and proofreading. H.R.\ and B.S.\ acknowledge support from the Agency for Innovation by Science and Technology in Flanders (IWT nr.\ 111504) and KU Leuven Research Fund (PF/2010/07). R.A.\ acknowledges support from Fundaci\'o ``La Caixa'', MINECO (FIS2013-41144-P), and DURSI (2014-SGR-878). I.P.\ acknowledges support from MINECO (FIS 2011-22603), DURSI (2014SGR-922), and Generalitat de Catalunya under Program Icrea Acad\`emia.
\end{acknowledgments}

\footnotesize{
\xfigtextfont{

}
}
%----------------------------------------------------------------------------------------

\end{article}

% 
% %----------------------------------------------------------------------------------------
% 
\end{document}